\documentclass[12pt]{article}
\usepackage{amsfonts}
\usepackage{amsmath}
\usepackage{epsfig}
\usepackage{graphicx}

\begin{document}

\title{\textbf{The fractional volatility model and rough volatility}}
\author{\textbf{R. Vilela Mendes}\thanks{%
rvilela.mendes@gmail.com; rvmendes@ciencias.ulisboa.pt;
https://label2.tecnico.ulisboa.pt/vilela/} \\
CMAFcIO, Faculdade de Ci\^{e}ncias, \\
Universidade de Lisboa}
\date{}
\maketitle

\begin{abstract}
The question of the volatility roughness is interpreted in the framework of
a data-reconstructed fractional volatility model, where volatility is driven
by fractional noise. Some examples are worked out and also, using Malliavin
calculus for fractional processes, an option pricing equation and its
solution are obtained.
\end{abstract}


\section{The fractional volatility model}

Some years ago, in collaboration with M. J. Oliveira \cite{dp2008-22}, a
program was started to reconstruct the market process from the data, using
only minimal mathematical and theoretical prejudices. Consistency with the
data was the main concern and only two general hypothesis were used, namely:

(\textbf{H1}) The log-price process $\log S_{t}$ belongs to a probability
product space $\Omega \otimes \Omega ^{^{\prime }}$ of which the first one, $%
\Omega $, is the Wiener space and the second, $\Omega ^{^{\prime }}$, is a
probability space to be reconstructed from the data. Denote by $\omega \in
\Omega $ and $\omega ^{^{\prime }}\in \Omega ^{^{\prime }}$ the elements
(sample paths) in $\Omega $ and $\Omega ^{^{\prime }}$ and by $\mathcal{F}%
_{t}$ and $\mathcal{F}_{t}^{^{\prime }}$ the $\sigma -$algebras in $\Omega $
and $\Omega ^{^{\prime }}$ generated by the processes up to $t$. Then, a
particular realization of the log-price process would be denoted 
\begin{equation*}
\log S_{t}\left( \omega ,\omega ^{^{\prime }}\right)
\end{equation*}%
This first hypothesis is really not limitative. Even if none of the
non-trivial stochastic features of the log-price were captured by Brownian
motion, that would simply mean that $S_{t}$ is a trivial function in $\Omega 
$.

(\textbf{H2}) The second hypothesis is stronger, although natural. It is
assumed that for each fixed $\omega ^{^{\prime }}$, $\log S_{t}\left(
\bullet ,\omega ^{^{\prime }}\right) $ is a square integrable random
variable in $\Omega $.

\begin{center}
---------
\end{center}

From the second hypothesis it follows that, for each fixed $\omega
^{^{\prime }}$, 
\begin{equation}
\begin{array}{lll}
\frac{dS_{t}}{S_{t}}\left( \bullet ,\omega ^{^{\prime }}\right) & = & \mu
_{t}\left( \bullet ,\omega ^{^{\prime }}\right) dt+\sigma _{t}\left( \bullet
,\omega ^{^{\prime }}\right) dB\left( t\right)%
\end{array}
\label{1.1}
\end{equation}%
where $\mu _{t}\left( \bullet ,\omega ^{^{\prime }}\right) $ and $\sigma
_{t}\left( \bullet ,\omega ^{^{\prime }}\right) $ are well-defined processes
in $\Omega $ (Theorem 1.1.3 in Ref.\cite{Nualart}). The process associated
to the probability space $\Omega ^{^{\prime }}$ was then inferred from the
data and this data-reconstructed $\sigma _{t}$ process was called the 
\textit{induced volatility}.

The scaling properties of the data-reconstructed induced volatility process
were then carefully analyzed \cite{dp2008-22}. The conclusion was that the
log-price, the volatility and the log-volatility are not self-similar
processes and it is only after the log-volatility is integrated and the
linear part extracted, that a self-similar process $R_{\sigma }\left(
t\right) $ is obtained. This is an essential finding of the
data-reconstructed model. That volatility is modeled by the finite
differences of a self-similar process is an essential difference from other
models that take into account the long-range correlation of the volatility.
In some models \cite{Comte} \cite{Gloter} \cite{Djehiche} it is fractional
Brownian motion (fBm) itself that drives the volatility, not a derivative of
this process.

The simplest mathematical processes having these properties were identified
and the following stochastic volatility model proposed

\begin{equation}
\begin{array}{lll}
dS_{t} & = & \mu S_{t}dt+\sigma _{t}S_{t}dB\left( t\right) \\ 
\log \sigma _{t} & = & \beta +\frac{k}{\delta }\left\{ B_{H}\left( t\right)
-B_{H}\left( t-\delta \right) \right\}%
\end{array}
\label{1.2}
\end{equation}%
\textit{This fractional volatility model (FVM) is the minimal model that is
consistent both with the mathematical hypothesis H1 and H2 and the scaling
properties of the data.}

$\delta $ is the observation time scale (one day, for daily data). In this
model the volatility is not driven by fractional Brownian motion but by
fractional noise. For the volatility (at resolution $\delta $) 
\begin{equation}
\sigma \left( t\right) =\theta e^{\frac{k}{\delta }\left\{ B_{H}\left(
t\right) -B_{H}\left( t-\delta \right) \right\} -\frac{1}{2}\left( \frac{k}{%
\delta }\right) ^{2}\delta ^{2H}}  \label{1.3}
\end{equation}%
the term $-\frac{1}{2}\left( \frac{k}{\delta }\right) ^{2}\delta ^{2H}$
insuring that $E\left( \sigma \left( t\right) \right) =\theta $. In (\ref%
{1.2}) the constant $k$ measures the strength of the volatility randomness.
In the $\delta \rightarrow 0$ limit the driving process would be the
distribution-valued process $W_{H}$%
\begin{equation}
W_{H}=\lim_{\delta \rightarrow 0}\frac{1}{\delta }\left( B_{H}\left(
t\right) -B_{H}\left( t-\delta \right) \right)  \label{1.4}
\end{equation}

Explicit expressions for the distribution of price returns were obtained 
\cite{dp2008-22} and one interesting feature was the fact that, once the
parameters were obtained from daily data in one market, then the model was
also consistent with high-frequency data in a different market by simply
changing the time scale $\delta $. This seemed somewhat mysterious until
comparison with agent based models \cite{Agent-based} revealed that in
business-as-usual days, the random fluctuations depend more on the limit
order book price mechanism than on the individual actions of the market
players.

Further study \cite{Arbitrage} revealed that the FVM model (\ref{1.2}) is
arbitrage free. Furthermore, when the two sources of randomness (in $B\left(
t\right) $ and $B_{H}\left( t\right) $) are related by an integral
representation for $B_{H}\left( t\right) $ the market is also complete%
\footnote{%
To associate the Brownian and fractional Brownian processes to the same
underlying probability space is quite natural in the framework of white
noise stochastic analysis \cite{Duncan} \cite{Biagini}.}.

\section{Rough volatility}

Some time ago Gatheral and collaborators \cite{Gatheral}, working in the
context of the Comte-Renault model \cite{Comte}, suggested, by the analysis
of the roughness of volatility data, a value $H<\frac{1}{2}$ for the Hurst
index. In a fractional Brownian motion process the Hurst index characterizes
both the roughness and the correlation or anticorrelation of the process.
Hence, if the volatility is driven by fractional Brownian motion, a Hurst
index smaller than $\frac{1}{2}$ would seem to contradict the market
long-range dependent volatility (volatility clustering) \cite{Crato} \cite%
{Cont1}. The way the realized volatility is measured at high frequency has
however been criticized by several authors \cite{Fukasawa} \cite{Rogers2} 
\cite{Cont2} suggesting that the origin of the roughness lies in the
microstructure noise \cite{Lahiri} rather than on the actual volatility
process.

However there is no real contradiction in the framework of the fractional
volatility model (\ref{1.2}) because there the volatility is driven by
fractional noise (fN), not by fractional Brownian motion (fBm). Fig.\ref%
{Noise} compares a simulated path of $10000$ steps of fBm at $H=0.8$ with
the corresponding one-step fractional noise ($B_{H}\left( t+1\right)
-B_{H}\left( t\right) $). 

\begin{figure}[htb]
\centering
\includegraphics[width=0.5\textwidth]{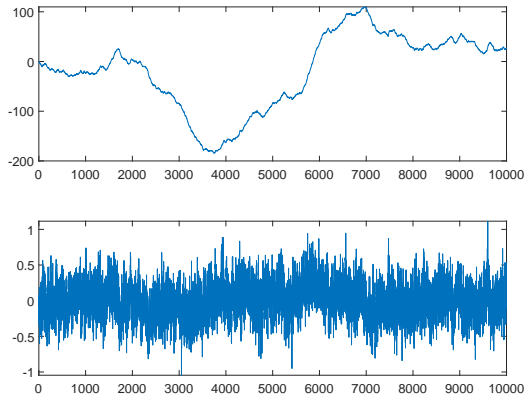}
\caption{Fractional Brownian motion at $H=0.8$ and the corresponding one-step
fractional noise}
\label{Noise}
\end{figure}

One sees that the apparent roughness of the fractional noise
mimics fBm at $H\simeq 0.1$. Therefore, using the hypothesis, as in Comte
and Renault \cite{Comte}, that it is fBm that drives volatility, one obtains
the wrong Hurst index. What the data analysis performed in \cite{dp2008-22}
implies is that, only when log-volatility is integrated and the linear part
extracted, is a self-similar process $R_{\sigma }\left( t\right) $ obtained.
Hence, long-range dependence and self-similarity are a property of
integrated log-volatility, not of volatility itself.

\begin{figure}[htb]
\centering
\includegraphics[width=0.5\textwidth]{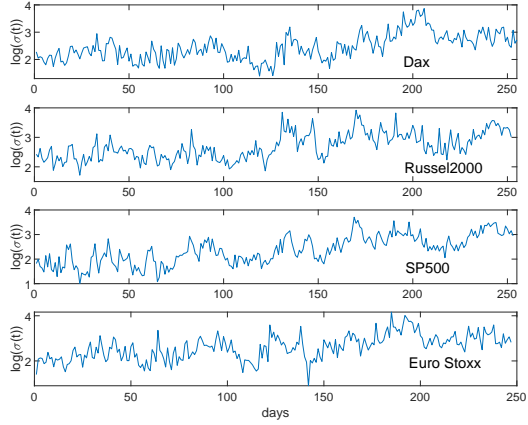}
\caption{One-day volatility data for 4 indexes (May 2021-May 2022)}
\label{indexes}
\end{figure}

As a check I have picked up some volatility data \cite{volat-data} and
performed the same analysis as in \cite{dp2008-22}. The data that was
analyzed was one-day volatility $\sigma \left( t\right) $ for the indexes
DAX, Russel 2000, S\&P500 and EURO STOXX 50 for the period 20/05/2021 to
22/05/2022 (Fig.\ref{indexes}), as well as 6 minute data for S\&P500 for the
period 24/11/2021 to 24/05/2022.

\begin{figure}[htb]
\centering
\includegraphics[width=0.5\textwidth]{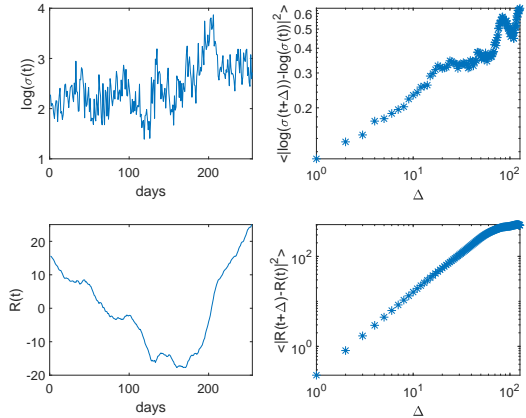}
\caption{The fractional volatility analysis for the DAX
index}
\label{dax_1d}
\end{figure}

Fig.\ref{dax_1d} displays the results for the DAX index.

\begin{figure}[htb]
\centering
\includegraphics[width=0.5\textwidth]{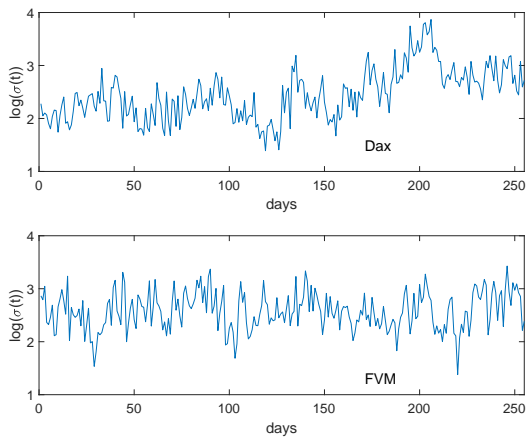}
\caption{DAX volatility versus a corresponding
fractional volatility model sample path}
\label{dax_vs_fvm}
\end{figure}

The upper left panel displays $\log
\sigma \left( t\right) $. Then, if this quantity were to follow fractional
Brownian motion, as some authors have assumed, one would expect%
\begin{equation*}
\mathbb{E}\left\{ \left( \log \sigma \left( t+\Delta \right) -\log \sigma
\left( t\right) \right) ^{2}\right\} \sim \Delta ^{2H}
\end{equation*}%
The upper right panel of Fig.\ref{dax_1d} clearly suggests that this is a
bad hypothesis. In the figure $\left\langle \bullet \right\rangle $ stands
for the empirical average, an empirical proxy for the expectation value.
Next, I have formed the integrated log-volatility and after the extraction
of the linear part $\beta t$ one obtains the process $R\left( t\right) $ (in
the lower left panel)%
\begin{equation*}
\sum_{n=0}^{t/\delta }\log \sigma \left( n\delta \right) =\beta t+R\left(
t\right)
\end{equation*}%
Computing $\left\langle \left( R\left( t+\Delta \right) -R\left( t\right)
\right) ^{2}\right\rangle \approx \mathbb{E}\left\{ \left( R\left( t+\Delta
\right) -R\left( t\right) \right) ^{2}\right\} $ one concludes (see the
lower right panel of Fig.\ref{dax_1d}) that%
\begin{equation*}
\mathbb{E}\left\{ \left( R\left( t+\Delta \right) -R\left( t\right) \right)
^{2}\right\} \sim \Delta ^{2H}
\end{equation*}%
and the identification of $R\left( t\right) $ with fractional Brownian
motion is a reasonable hypothesis. Hence%
\begin{equation*}
\log \sigma \left( t\right) =\beta +\frac{k}{\delta }\left\{ B_{H}\left(
t\right) -B_{H}\left( t-\delta \right) \right\}
\end{equation*}%
The following table shows the values of $\beta $ and $H$ that are obtained
for the indexes that were analyzed%
\begin{equation*}
\begin{tabular}{|l|l|l|}
\hline
& H & $\beta $ \\ \hline
SP500\_1d & 0.85 & 2.35 \\ \hline
SP500\_6min & 0.86 & 2.77 \\ \hline
Russel2000\_1d & 0.84 & 2.68 \\ \hline
Euro\_Stoxx\_1d & 0.84 & 2.59 \\ \hline
DAX\_1d & 0.88 & 2.40 \\ \hline
\end{tabular}%
\end{equation*}%
Fig.\ref{dax_vs_fvm} compares the DAX volatility data with a simulated
sample path of the fractional volatility model (FVM) with the $H$ and $\beta 
$ values listed in the table. The model, having the same statistical
properties as the data, might to be said to provide a "perfect simulation" 
\cite{Galves} of the data. However it must be pointed out, that perfect
simulation in the statistical sense only means the same statistical
properties, it is not "perfect forecasting". An attempt has been made in the
past \cite{Nuno} to use the FVM to forecast volatility. The main conclusion
was that although having the good statistical properties, FVM was not
optimal as a forecasting device, because the market seemed to have, in
addition to the stochastic terms in (\ref{1.2}), a deterministic
mean-reverting component, for example%
\begin{equation*}
\log \sigma \left( t\right) =\beta e^{\alpha \left( \beta -\log \sigma
\left( t\right) \right) t}+\frac{k}{\delta }\left\{ B_{H}\left( t\right)
-B_{H}\left( t-\delta \right) \right\}
\end{equation*}%
Such mean-reverting effect is in fact suggested by a close examination of
the data plots and their comparison with the sample paths of the FVM. 

Finally, to the question of whether volatility is rough, the answer is yes,
but only because it is driven by fractional noise with $H>\frac{1}{2}$.

One of the motivations for the development of reliable volatility models
lies on the problem of option pricing and on corrections to Black-Scholes.
In particular the rough volatility framework has also been used for option
pricing \cite{Gatheral-option}. In \cite{dp2008-22}, using a simple-minded
extension of the Hull-White \cite{Hull} reasoning, an approximate formula
was obtained with features similar to those observed in options data. Here a
more rigorous derivation will be made, based on the mathematical framework
that has been developed for the stochastic analysis of fractional processes.

\section{Fractional stochastic analysis and option pricing}

The idea of using fractional processes as a tool for modeling in Finance
appeared at an early stage \cite{Mandelbrot2}. However, because it was
pointed out \cite{Rogers} that markets based on $B_{H}\left( t\right) $
could have arbitrage, fractional Brownian motion (fBm) was not considered,
for a while, as a promising tool for mathematical modeling in Finance. The
arbitrage result in \cite{Rogers} is a consequence of using pathwise
integration. With a different definition \cite{Duncan}, 
\begin{equation*}
\int_{a}^{b}f\left( t,\omega \right) dB_{H}\left( t\right) =\lim_{\left\vert
\Delta \right\vert \rightarrow 0}\sum_{k}f\left( t_{k},\omega \right)
\diamond \left( B_{H}\left( t_{k+1}\right) -B_{H}\left( t_{k}\right) \right)
\end{equation*}%
where $\Delta :a=t_{0}<t_{1}<\cdots <t_{n}=b$ is a partition of the interval 
$[a,b]$, $\left\vert \Delta \right\vert =\max_{0\leq k\leq n-1}\left(
t_{k+1}-t_{k}\right) $ and $\diamond $ denotes the Wick product, the
integral has zero expectation value and the arbitrage result is no longer
true. This is, in fact, the most natural definition because it is the Wick
product that is associated to integrals of It\^{o} type, whereas the usual
product is natural for integrals of Stratonovich type. An essentially
equivalent approach constructs the stochastic integral through the
divergence operator and Malliavin calculus \cite{Nualart2}. In any case if
fBm is included in the log return process \cite{Oksendal2} \cite{Elliott2004}
it contradicts the empirical short autocorrelation of this process. The
conclusion is that fractional processes are only relevant to drive the
volatility, not the log-price itself.

A fully consistent stochastic calculus has now been developed for fractional
Brownian motion \cite{Duncan} \cite{Biagini} \cite{Nualart2} \cite{Oksendal2}%
\ \cite{Ustunel} \cite{Hu} \cite{Oksendal1} \cite{Alos} and this is the
setting that will be used here to derive an option pricing equation.

Because volatility is not a tradable security, a pure arbitrage argument
cannot completely determine the fair price of an option. On the other hand,
because of the fractional nature of the volatility process, volatility
follows a stochastic process different from the one of the underlying
security. Therefore, we cannot apply the reasoning \cite{Lyuu} that leads to
uniform coefficients of the form $\left( \mu _{i}-\lambda _{i}\sigma
_{i}\right) $ in the first derivative terms of the option pricing equation%
\footnote{$\mu _{i}$, $\sigma _{i}$ and $\lambda _{i}$ would be the drift,
volatility and market price of risk for each process}. Hence, a first
principles derivation, with clearly specified assumptions is required.

As in Black-Scholes \cite{Black} \cite{Merton} form a portfolio 
\begin{equation}
\Pi \left( t\right) =V\left( S,\sigma ,t\right) -\Delta \left( S,\sigma
,t\right) S_{t}  \label{4.1}
\end{equation}%
To compute the stochastic differential of $\Pi \left( t\right) $ one uses
the It\^{o} formula for the price process ($S_{t}$) and the fractional It%
\^{o} formula \cite{Duncan} Namely, if $dX_{t}=c\left( t,\omega \right)
dB_{H}\left( t\right) $ , then 
\begin{equation*}
df\left( t,X_{t}\right) =\frac{\partial f}{\partial t}dt+\frac{\partial f}{%
\partial X}dX_{t}+\frac{\partial ^{2}f}{\partial X^{2}}c\left( t,\omega
\right) D_{t}^{\phi }\left( X_{t}\right)
\end{equation*}%
$D_{t}^{\phi }\left( X_{t}\right) $ being the $\phi -$Malliavin derivative
corresponding to the $X_{t}-$ process, defined by 
\begin{eqnarray*}
D_{f}^{\phi }X_{t}\left( \omega \right) &=&\lim_{\varepsilon \rightarrow 0}%
\frac{1}{\varepsilon }\left\{ X\left( \omega +\varepsilon \int_{0}^{\bullet
}ds\int_{0}^{\infty }\phi \left( s,u\right) f\left( u\right) du\right)
-X\left( \omega \right) \right\} \\
&=&\int_{0}^{\infty }D_{u}^{\phi }\left( X_{t}\right) f\left( u\right) du
\end{eqnarray*}%
$\phi \left( s,u\right) $ being the kernel 
\begin{equation*}
\phi \left( s,u\right) =H\left( 2H-1\right) \left\vert s-u\right\vert ^{2H-2}
\end{equation*}%
for $\frac{1}{2}<H<1$

Then from (\ref{1.2}), choosing $\Delta \left( S,\sigma ,t\right) =\frac{%
\partial V}{\partial S}$ one obtains 
\begin{eqnarray*}
d\Pi \left( t\right) &=&\left\{ \frac{\partial V}{\partial t}+\frac{1}{2}%
\frac{\partial ^{2}V}{\partial S^{2}}\sigma ^{2}S^{2}\right\} dt+\frac{%
\partial V}{\partial \left( \log \sigma \right) }\frac{k}{\delta }\left(
dB_{H}\left( t\right) -dB_{H}\left( t-\delta \right) \right) \\
&&+\frac{\partial ^{2}V}{\partial \left( \log \sigma \right) ^{2}}\frac{k^{2}%
}{\delta ^{2}}D_{t}^{\phi }\left( B_{H}\left( t\right) -B_{H}\left( t-\delta
\right) \right)
\end{eqnarray*}%
and computing the Malliavin derivative 
\begin{eqnarray}
d\Pi \left( t\right) &=&\left\{ \frac{\partial V}{\partial t}+\frac{1}{2}%
\frac{\partial ^{2}V}{\partial S^{2}}\sigma ^{2}S^{2}\right\} dt+\sigma 
\frac{\partial V}{\partial \sigma }\frac{k}{\delta }\left( dB_{H}\left(
t\right) -dB_{H}\left( t-\delta \right) \right)  \notag \\
&&+\left( \sigma ^{2}\frac{\partial ^{2}V}{\partial \sigma ^{2}}+\sigma 
\frac{\partial V}{\partial \sigma }\right) \frac{k^{2}}{\delta ^{2}}H\delta
^{2H-1}dt  \label{4.2}
\end{eqnarray}

In (\ref{4.2}) one is still left with the stochastic term $\sigma \frac{%
\partial V}{\partial \sigma }\frac{k}{\delta }\left( dB_{H}\left( t\right)
-dB_{H}\left( t-\delta \right) \right) $ and, because volatility is not a
tradable security this term cannot be eliminated by a portfolio choice.
Instead one identifies this term with a (deterministic) market price of
volatility term $\nu \frac{k}{\delta }\sigma \frac{\partial V}{\partial
\sigma }dt$ with $\nu >0$. Finally from $d\Pi \left( t\right) =r\Pi \left(
t\right) dt$, $r$ being the risk-free return, one ends up with 
\begin{equation}
\frac{\partial V}{\partial t}+rS\frac{\partial V}{\partial S}+\frac{\sigma
^{2}S^{2}}{2}\frac{\partial ^{2}V}{\partial S^{2}}+\frac{k}{\delta }\left(
kH\delta ^{2H-2}-\nu \right) \sigma \frac{\partial V}{\partial \sigma }%
+Hk^{2}\delta ^{2H-3}\sigma ^{2}\frac{\partial ^{2}V}{\partial \sigma ^{2}}%
=rV  \label{4.8}
\end{equation}%
as the general form of an option pricing equation consistent with the
stochastic volatility model in (\ref{1.2}).

One now obtains an integral representation for the solution of this equation
with the change of variable 
\begin{equation}
x=\log \frac{S}{K}  \label{4.9}
\end{equation}%
and passing to the two-dimensional Fourier transform 
\begin{equation}
V\left( t,x,\sigma \right) =\int \int d\phi d\rho F\left( \phi ,\rho ,\sigma
\right) e^{i\left( \phi t+\rho x\right) }  \label{4.10}
\end{equation}%
Then 
\begin{equation}
Hk^{2}\delta ^{2H-3}\sigma ^{2}\frac{\partial ^{2}F}{\partial \sigma ^{2}}+%
\frac{k}{\delta }\left( kH\delta ^{2H-2}-\nu \right) \sigma \frac{\partial F%
}{\partial \sigma }+\left( i\left( \phi +\rho r-\frac{\sigma ^{2}\rho }{2}%
\right) -\frac{\sigma ^{2}\rho ^{2}}{2}-r\right) F=0  \label{4.11}
\end{equation}

Defining new constants 
\begin{equation}
\begin{array}{lll}
\chi \left( \rho \right) & = & \frac{\nu }{2Hk\delta ^{2H-2}} \\ 
\xi ^{2}\left( \rho ,\phi \right) & = & \chi ^{2}\left( \rho \right) -\frac{%
r-i\left( \phi +\rho r\right) }{Hk^{2}\delta ^{2H-3}} \\ 
\zeta ^{2}\left( \rho \right) & = & -\frac{i\rho +\rho ^{2}}{2Hk^{2}\delta
^{2H-3}}%
\end{array}
\label{4.12}
\end{equation}%
and making the replacement 
\begin{equation}
F\left( \sigma \right) =\sigma ^{\chi }Z_{\xi }\left( \zeta \sigma \right)
\label{4.13}
\end{equation}%
Eq.(\ref{4.11}) reduces to a standard Bessel equation. Therefore the
solution of (\ref{4.8}) is 
\begin{equation}
V\left( t,x,\sigma \right) =\int \int d\rho d\phi e^{i\left( \phi t+\rho
x\right) }\sigma ^{\chi \left( \rho \right) }Z_{\xi (\rho ,\phi )}\left(
\zeta \left( \rho \right) \sigma \right)  \label{4.14}
\end{equation}%
$Z_{\xi }\left( \zeta \sigma \right) $ being a Bessel function. The Bessel
function will be a linear combination 
\begin{equation*}
Z_{\xi }\left( \zeta \sigma \right) =c_{1}J_{\xi }\left( \zeta \sigma
\right) +c_{2}N_{\xi }\left( \zeta \sigma \right)
\end{equation*}%
of a Bessel function of first kind and a Neumann function, with coefficients 
$c_{1}$ and $c_{2}$ to be fixed by the boundary condition, which for call
options is 
\begin{equation*}
V\left( T,x,\sigma \right) =\max \left( x,0\right)
\end{equation*}%
Eq.(\ref{4.14}) is an exact solution of the option pricing equation.

\end{document}